\documentclass[aps,pra,english,twocolumn,showpacs,preprintnumbers,amsmath,amssymb,floatfix]{revtex4-1}

\usepackage[T1]{fontenc}
\usepackage[latin9]{inputenc}
\usepackage{graphicx}
\usepackage{amssymb}
\usepackage{babel}


\usepackage{babel}
\makeatother

\begin{document}

\title{Casimir-Lifshitz interaction between ZnO and SiO$_2$ nanorods in bromobenzene: retardation effects turn  the interaction repulsive  at intermediate separations }

\author{Mathias Bostr{\"o}m}

\author{Bo E. Sernelius}
\email{bos@ifm.liu.se}
\affiliation{Division of Theory and Modeling, Department of Physics, 
Chemistry
and Biology, Link\"{o}ping University, SE-581 83 Link\"{o}ping, Sweden}

\author{Gustavo Baldissera}
\author{Clas Persson}
\altaffiliation{Department of Physics, University of Oslo, P. Box 1048 Blindern,
NO-0316 Oslo, Norway}
\affiliation{Dept of Materials Science and Engineering, Royal Institute of Technology, SE-100 44 Stockholm, Sweden}

\author{Barry W. Ninham}
\affiliation{Department of Applied Mathematics, Australian National University, Canberra, Australia}

\begin{abstract}
We consider the interaction between  a ZnO nanorod and a SiO$_2$ nanorod in bromobenzene. Using optical data for the interacting objects and ambient we calculate the force --- from short-range attractive van der Waals force to  intermediate range repulsive Casimir-Lifshitz force to long range entropically driven attraction.  The nonretarded van der Waals interaction is attractive at all separations. We demonstrate a retardation driven repulsion at intermediate separations. At short separations (in the nonretarded limit) and at large separations (in the classical limit) the interaction is attractive.  These effects can be understood from an analysis of multiple crossings of the dielectric functions of the three media as functions of imaginary frequencies.
\end{abstract}


\maketitle

Casimir predicted already in 1948\,\cite{Casi} the attraction between a pair of parallel, closely spaced, perfect conductors a distance apart.
Almost 50 years later Lamoreaux\,\cite{Lamo,Sush,Milt2} performed the first high accuracy measurement of Casimir forces between metal surfaces in vacuum. 	An interesting aspect of the Lifshitz-Casimir force is that according to theory it can also be repulsive\,\cite{Dzya,Rich1,Rich}.  The aim of this brief report is to demonstrate nonmonotonic distance dependence in the Lifshitz interaction between nanorods in solution. The force may go from attractive to repulsive to attractive as the separation is increased. Early theoretical and experimental efforts indirectly demonstrating repulsive Casimir-Lifshitz forces were reviewed in a recent communication \,\cite{BostPRA}.  Munday, Capasso, and Parsegian\,\cite{Mund} performed the first direct force measurements that demonstrated that the Lifshitz-Casimir force could be repulsive by a suitable choice of interacting surfaces in a fluid. They speculated that this effect could allow quantum levitation of nano-scale devices\,\cite{Mund}. 

When two objects are brought together correlations in  fluctuations of the charge and current densities in the objects or fluctuations of the fields usually result in an attractive force\,\cite{ Lond, Casi,Dzya,Lang, Maha, Isra, Ser, Milt, Pars, Ninhb}. At short distances this is the van der Waals force\,\cite{Lond}; at large distances the finite velocity of light becomes important (retardation effects) and the result is the Casimir force\,\cite{Casi, Milt}. With retardation effects we here mean all effects that appear because of the finite speed of light; not just the reduction in correlation between the charge density fluctuations at large distances; new propagating solutions to Maxwell's equations appear that bring correlations between current density fluctuations. When these propagating modes dominate we call the force Casimir force; when the surface modes\,\cite{Ser} dominate we call the force van der Waals force. The classical literature says that retardation is due to the finite speed of light that weakens the correlations. This is misleading. It is due to the quantum nature of light\,\cite{Wenn,Ninha}; this becomes obvious at the large separation limit where the weakening is independent of the speed of light.

Much interest has been focused on nanowires and nanorods \,\cite{Appell,Samuelson} and their potential applications in nanotechnology.  For instance ZnO  \,\cite{Yi,Zhao} and SiO$_2$  \,\cite{Nakamura} nanorods are interesting as potential building blocks in future nanomechanics and in life science applications \,\cite{Cui}.  Noruzifar et al. \,\cite{Noruzifar}, and much earlier Davies et al. \,\cite{Davies}, investigated Lifshitz forces between metal wires. 
The interaction between conducting particles can be surprisingly long range \,\cite{Davies}.
The interaction between rod-like particles was investigated by different groups \,\cite{Langbein1,Langbein2,Parsegian1972,Israelachvili1973,Mitchell1973,Lombardo,Mazzitelli}. Ninham and co-workers in particular investigated in great detail the Lifshitz-Casimir interaction between thin dielectric rods \,\cite{MitchCyl, Maha}. We will exploit their result  \,\cite{MitchCyl} obtained  for thin rods with diameters much smaller than the rod separation. We present calculations for a system with a  silica (SiO$_2$) nanorod  (2 nm diameter)  \,\cite{Grab} interacting with a zinc oxide (ZnO) nanorod (2 nm diameter)  across Bromobenzene (Bb)\,\cite{Mund}) and demonstrate that attractive and repulsive Lifshitz forces are produced in one and the same system. 

\begin{figure}
\includegraphics[width=7.55cm]{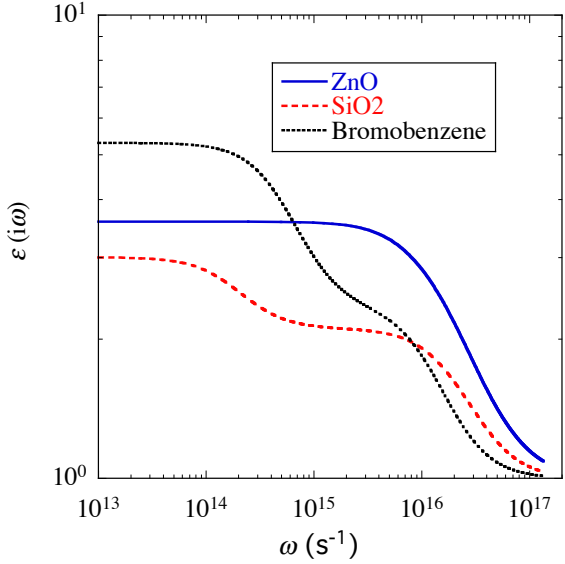}
\caption{(Color online) The dielectric function at imaginary frequencies for SiO$_2$ (silica)\,\cite{Grab}, Bb (bromobenzene)\,\cite{Mund}, and ZnO (zinc oxide).}
\label{figu1}
\end{figure}

The Casimir force can be calculated if the dielectric functions (for imaginary frequencies, $i\omega$) are known. These functions are shown in Fig.\,\ref{figu1}.  The dielectric function of ZnO on the imaginary frequency axis was obtained using a modified\,\cite{Ser} Kramers-Kronig dispersion relation applied to the function on the real axis. The function on the real axis was calculated using a partially self-consistent
GW approach\,\cite{ Kresse,Meira}, where the Green's function method was based on the
local density approximation within the density functional theory. There is a crossing between the curves for ZnO and Bb at a low frequency and another crossing between the curves for SiO$_2$ and Bb at a higher frequency. This opens up for the possibility of a transition of the interaction, from attraction to repulsion and back to attraction.

The interaction of material 1 (SiO$_2$) with material 2 (ZnO) across medium 3 (Bb) results in a summation of imaginary frequency terms\,\cite{Maha,MitchCyl}
\begin{equation}
F = \sum\limits_{n = 0}^\infty  {'g\left( {{\omega _n}} \right)} ;\quad {\omega _n} = \frac{{2\pi n}}{{\hbar \beta }},
\label{equ1}
\end{equation}
where
\begin{equation}
g\left( {{\omega _n}} \right) = \frac{1}{\beta }\int\limits_0^\infty  {\frac{{dk}}{{2\pi }}G\left( {{\omega _n},k} \right)},
\label{equ2}
\end{equation}
and
\begin{equation}
\begin{array}{*{20}{l}}
G\left( {{\omega _n},k} \right) =  - \frac{{{a^2}{b^2}}}{4}\frac{{\left( {{\varepsilon _3} - {\varepsilon _1}} \right)\left( {{\varepsilon _3} - {\varepsilon _2}} \right)}}{{\varepsilon _3^2}}\gamma _3^4K_0^2\left( {{\gamma _3}R} \right)\\
\quad \quad  - \frac{{{a^2}{b^2}}}{2}\left\{ {\left( {\frac{{{\varepsilon _3}}}{{\gamma _3^2}} - \frac{{{\varepsilon _2}}}{{\gamma _2^2}}} \right)\frac{{\left( {{\varepsilon _3} - {\varepsilon _1}} \right)}}{{{\varepsilon _3}\left( {{\varepsilon _3} + {\varepsilon _2}} \right)}}} \right.\\
\quad \quad \quad \quad  + \left. {\left( {\frac{{{\varepsilon _3}}}{{\gamma _3^2}} - \frac{{{\varepsilon _1}}}{{\gamma _1^2}}} \right)\frac{{\left( {{\varepsilon _3} - {\varepsilon _2}} \right)}}{{{\varepsilon _3}\left( {{\varepsilon _3} + {\varepsilon _1}} \right)}}} \right\}\gamma _3^6K_1^2\left( {{\gamma _3}R} \right)\\
 - \frac{{{a^2}{b^2}}}{2}\left\{ {\left( {\frac{{{\varepsilon _3}}}{{\gamma _3^2}} - \frac{{{\varepsilon _1}}}{{\gamma _1^2}}} \right)\left( {\frac{{{\varepsilon _3}}}{{\gamma _3^2}} - \frac{{{\varepsilon _2}}}{{\gamma _2^2}}} \right)\frac{{\gamma _3^4}}{{\left( {{\varepsilon _3} + {\varepsilon _2}} \right)\left( {{\varepsilon _3} + {\varepsilon _1}} \right)}}} \right.\\
\quad \quad  + \frac{{\varepsilon _3^2\gamma _1^2\gamma _2^2}}{{\left( {{\varepsilon _3} + {\varepsilon _2}} \right)\left( {{\varepsilon _3} + {\varepsilon _1}} \right)}}\left( {\frac{1}{{\gamma _3^2}} - \frac{1}{{\gamma _1^2}}} \right)\left( {\frac{1}{{\gamma _3^2}} - \frac{1}{{\gamma _2^2}}} \right)\\
\quad \quad  + \left( {\frac{{{\varepsilon _3}}}{{\gamma _3^2}} - \frac{{{\varepsilon _1}}}{{\gamma _1^2}}} \right)\left( {\frac{1}{{\gamma _3^2}} - \frac{1}{{\gamma _2^2}}} \right)\frac{{\varepsilon _3^2\gamma _1^2\gamma _2^2}}{{\left( {{\varepsilon _3} + {\varepsilon _2}} \right)\left( {{\varepsilon _3} + {\varepsilon _1}} \right)}}\\
\quad \quad  + \left. {\left( {\frac{{{\varepsilon _3}}}{{\gamma _3^2}} - \frac{{{\varepsilon _2}}}{{\gamma _2^2}}} \right)\left( {\frac{1}{{\gamma _3^2}} - \frac{1}{{\gamma _1^2}}} \right)\frac{{\varepsilon _3^2\gamma _1^2\gamma _2^2}}{{\left( {{\varepsilon _3} + {\varepsilon _2}} \right)\left( {{\varepsilon _3} + {\varepsilon _1}} \right)}}} \right\}\\
\quad \quad \quad \quad \quad \quad \quad  \times \left\{ {\gamma _3^4\left[ {K_2^2\left( {{\gamma _3}R} \right) + K_0^2\left( {{\gamma _3}R} \right)} \right]} \right\}.
\end{array}
\label{equ3}
\end{equation}
\begin{figure}
\includegraphics[width=7.55cm]{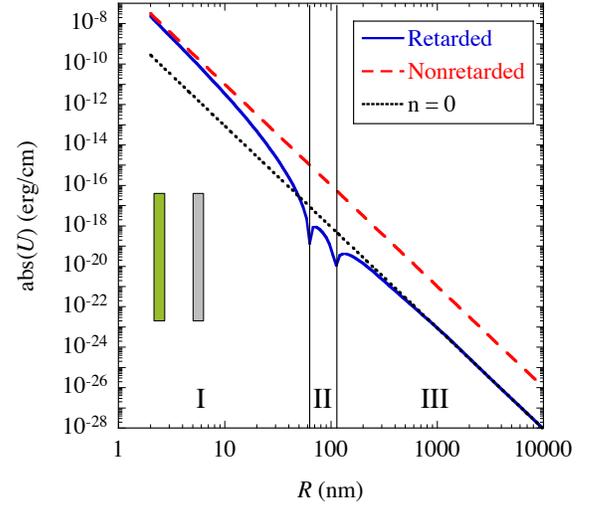}
\caption{(Color online) The Casimir-Lifshitz interaction energy per unit length between ZnO and SiO$_2$ nanorods (each with 2 nm diameter) in bromobenzene. The fully retarded free energy is attractive in regions I and III but turns repulsive in region II. The nonretarded energy and the n=0 contribution are both attractive and they decay $\propto R^{-5}$.}
\label{figu2}
\end{figure}

The momentum dependence is found in the functions ${\gamma _\alpha } = \sqrt {{k^2} + {\varepsilon _\alpha }{{\left( {{\omega _n}/c} \right)}^2}} $. The parameters $a$ and $b$ are the radii of the two rods. Note that positive values of $F$ correspond to repulsive interaction. 
We recently investigated the interaction between a planar gold surface interacting with a planar silica surface in bromobenzene \,\cite{BostPRA}. There the nonretarded van der Waals interaction was attractive while  both the long-range Casimir asymptote and the long-range entropic asymptote were found to be repulsive. In contrast in the present system shown in Fig.\,\ref{figu2}  it is only for intermediate separations that retardation turns the interaction repulsive. The reversal of the sign of the Casimir energy with retardation for intermediate separations as compared to without retardation is due to a subtle balance of attractive (high and low frequencies) and repulsive (intermediate frequencies) contributions. When the separation increases from its lowest value the attractive contributions at high frequencies are weakened by retardation at a higher rate than  the repulsive contributions at intermediate frequencies. At large separations the low frequencies dominate giving again attraction. The distance-dependent Bessel functions, ${K_i}\left( {R{\gamma _3}} \right) = {K_i}\left[ {R\sqrt {{k^2} + {\varepsilon _{Bb}}\left( {i{\omega _n}} \right){{\left( {{\omega _n}/c} \right)}^2}} } \right]$, in the Lifshitz expression for the Casimir energy, Eq.\,(\ref{equ3}), has the effect that all frequencies contribute to the nonretarded van der Waals force while only low frequencies contribute in the long range retarded Casimir regime. The ultimate long-range asymptote, the entropic term, only includes the zero frequency mode.  The zero frequency contribution is attractive and follows the same power law as the nonretarded asymptote.  It is seen to be possible to via the optical properties tune the materials used such that they give repulsion at specific separations. In this case only at intermediate separations. There is also a, to our knowledge not yet experimentally observed, maximum of both the repulsive and attractive parts of the Lifshitz-Casimir interaction energy.

\begin{figure}
\includegraphics[width=7.55cm]{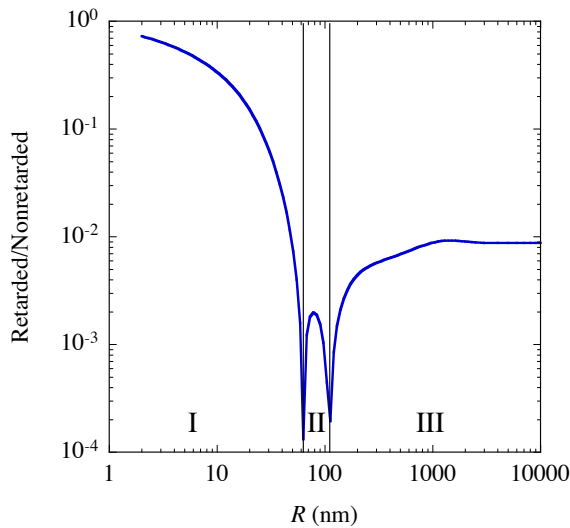}
\caption{(Color online) The absolute value of the ratio between the retarded and nonretarded interaction free energy per unit length (the nonretarded energy is attractive and  $\propto R^{-5}$). The fully retarded free energy is attractive in regions I and III but turns repulsive in region II.}
\label{figu3}
\end{figure}

We show  in Fig.\,\ref{figu3} the absolute value of the ratio between the retarded and nonretarded interaction free energy. The aim of this figure is to illustrate  the deviations from the simple attractive nonretarded  $R^{-5}$-power law. It is clear that retardation changes quite dramatically the interaction between silica and zinc oxide nanorods.

Using optical measurements on the interacting objects and liquid makes it possible to predict the force --- from short-range attractive van der Waals force to  intermediate range repulsive Casimir force to long range entropically driven attraction. It should be noted that at 1 $\mu$m separation the thermal entropic effect starts to become important. This is similar to what we found for the Casimir force between real metal surfaces in vacuum\,\cite{Bost2} and between gold and silica in bromobenzene \,\cite{BostPRA}.

\begin{acknowledgments}
The research was sponsored by the VR-contract No:70529001 which is gratefully acknowledged.
\end{acknowledgments}


\begin{thebibliography}{10}
\bibitem{Casi} H. B. G. Casimir, Proc. K. Ned. Akad. Wet. {\bf 51}, 793 (1948). 
\bibitem{Lamo}S. K. Lamoreaux, Phys. Rev. Lett. {\bf 78}, 5 (1997); {\bf 81}, 5475 (1998).
\bibitem{Sush} A. O. Sushkov, W. J. Kim, D. A. R. Dalvit, and S. K. Lamoreaux, Nature Physics, {\bf 7}, 230, 2011.
\bibitem{Milt2} K. Milton, Nature Physics, {\bf 7}, 190, 2011.
\bibitem{Dzya} I. E. Dzyaloshinskii, E. M. Lifshitz, and P. P. Pitaevskii, Advan. Phys. {\bf 10}, 165 (1961).
\bibitem{Rich1} P. Richmond and B. W. Ninham, Solid State Communications {\bf 9}, 1045 (1971).
\bibitem{Rich} P. Richmond, B. W. Ninham and R. H. Ottewill, J. Colloid Int. Sci. {\bf 45}, 69 (1973).
\bibitem{BostPRA} M. Bostr{\"o}m, B. E. Sernelius, I. Brevik, and B. W. Ninham, Phys. Rev. A. {\bf 85}, 010701 (2012).
\bibitem{Mund} J. N. Munday, F. Capasso, and V. A. Parsegian, Nature, {\bf 457}, 07610 (2009).
\bibitem{Lang}  D. Langbein, {\it Theory of Van der Waals attraction}, (Springer, New York, 1974,) in the series Springer Tracts in Modern Physics.
\bibitem{Lond} F. London, Z. Phys. Chem. B {\bf 11}, 222 (1930).
\bibitem{Maha} J. Mahanty and B. W. Ninham, {\it Dispersion Forces}, (Academic Press, London, 1976).
\bibitem{Isra} J. Israelachvili, {\it Intermolecular and Surface Forces}, 2nd Ed., (Academic Press, London, 1991).
\bibitem{Ser} Bo E. Sernelius, {\it Surface Modes in Physics} (Wiley-VCH,  Berlin, 2001).
\bibitem{Milt} K. A. Milton, {\it The Casimir Effect: Physical Manifestations of Zero-Point Energy}, (World Scientific, Singapore, 2001).
\bibitem{Pars} V. A. Parsegian, {\it Van der Waals forces: A handbook for biologists, chemists, engineers, and physicists}, (Cambridge University Press, New York, 2006). 
\bibitem{Ninhb} B. W. Ninham and P. Lo Nostro, {\it Molecular Forces and Self Assembly: in Colloid, Nano Sciences and Biology}, (Cambridge University Press, Cambridge, 2010).
\bibitem{Wenn} H. Wennerstr{\"o}m, J. Daicic, and B. W. Ninham, Phys. Rev. {\bf A}  60, 2581 (1999).
\bibitem{Ninha} B. W. Ninham, and J. Daicic, Phys. Rev. {\bf A}  57, 1870 (1998).
\bibitem{Appell} D. Appell, Nature {\bf 419}, 553 (2002).
\bibitem{Samuelson} L. Samuelson, Mater. Today {\bf 6}, 22 (2003).
\bibitem{Yi} G.-C. Yi, C. Wang, and W. Il Park, Semicond. Sci. Technol. {\bf 20}, 522 (2005).
\bibitem{Zhao}Q.X. Zhao, L.L. Yang, M. Willander, Bo E. Sernelius, and  P.O. Holtz, J. Appl. Phys. {\bf 104}, 073526 (2008). 
\bibitem{Nakamura} H. Nakamura and Y. Matsui, J. Am. Chem. Soc. {\bf 151}, 2651 (1995).
\bibitem{Cui} Y. Cui, Q. Wei, H. Park, and C. M. Lieber, Science {\bf 293}, 1289 (2001).
\bibitem{Noruzifar} E. Noruzifar, T. Emig, and R. Zandi, Phys. Rev. A. {\bf 84}, 042501 (2011)
\bibitem{Davies} B. Davies, B. W. Ninham, and P. Richmond, J. Chem. Phys. {\bf 58}, 744 (1973).
\bibitem{Langbein1} D. Langbein, Phys. Kondens. Materie {\bf 15}, 61 (1972).
\bibitem{Langbein2} D. Langbein, Adv. Solid State Phys. (Festk{\"o}rperprobleme) {\bf XIII}, 85 (1973).
\bibitem{Parsegian1972} V. A. Parsegian, J. Chem. Phys. {\bf 56}, 4393 (1972).
\bibitem{Israelachvili1973} J. N. Israelachvili, J. Theoret. Biol. {\bf 42}, 411 (1973).
\bibitem{Mitchell1973} D. J. Mitchell and B. W. Ninham, J. Chem. Phys. {\bf 59}, 1246 (1973).
\bibitem{Lombardo} F. C. Lombardo, F. D. Mazzitelli,  P. I. Villar, and D. A. R. Dalvit, Phys. Rev. A {\bf 82}, 042509 (2010).
\bibitem{Mazzitelli} F. D. Mazzitelli, D. A. R. Dalvit, and F. C. Lombardo, New Journal of Physics {\bf 8}, 240 (2006).
\bibitem{MitchCyl} D. J. Mitchell, B. W. Ninham, and P. Richmond, Biophys. J. {\bf 13}, 370 (1973).
 \bibitem{Grab} A. Grabbe, Langmuir {\bf 9}, 797 (1993).
\bibitem{Kresse} G. Kresse and D. Joubert, Phys. Rev. B 59, 1758 (1999).
\bibitem{Meira} M. V. Castro Meira, A. Ferreira da Silva, G. Baldissera, C. Persson, J. A. Freitas, Jr., N. Gutman, A. Saar, O. Nur, and M. Willander, submitted.
\bibitem{Bost2} M. Bostr{\"o}m and B. E. Sernelius, Phys. Rev. Lett. {\bf 84}, 4757 (2000).

 
\end{thebibliography}
\end{document}